\documentclass[twocolumn,reprint,amsmath,amssymb,aps,superscriptaddress,prl]{revtex4}
\usepackage[latin9]{inputenc}
\usepackage{color}
\usepackage{float}
\usepackage{units}
\usepackage{amsmath}
\usepackage{graphicx}

\makeatletter
\@ifundefined{textcolor}{}
{%
 \definecolor{BLACK}{gray}{0}
 \definecolor{WHITE}{gray}{1}
 \definecolor{RED}{rgb}{1,0,0}
 \definecolor{GREEN}{rgb}{0,1,0}
 \definecolor{BLUE}{rgb}{0,0,1}
 \definecolor{CYAN}{cmyk}{1,0,0,0}
 \definecolor{MAGENTA}{cmyk}{0,1,0,0}
 \definecolor{YELLOW}{cmyk}{0,0,1,0}
}

\usepackage{dcolumn}
\usepackage{bm}
\usepackage{hyperref}
\usepackage{cleveref}

\makeatother

\begin{document}

\title{\textcolor{black}{Phase Stochastic Resonance in a forced nano-electromechanical
oscillator}}

\author{\textcolor{black}{Avishek Chowdhury}}

\affiliation{\textcolor{black}{Centre de Nanosciences et de Nanotechnologies,
CNRS, Univ. Paris-Sud, Universit� Paris-Saclay, C2N Marcoussis, 91460
Marcoussis, France}}

\author{\textcolor{black}{Sylvain Barbay}}

\affiliation{\textcolor{black}{Centre de Nanosciences et de Nanotechnologies,
CNRS, Univ. Paris-Sud, Universit� Paris-Saclay, C2N Marcoussis, 91460
Marcoussis, France}}

\author{\textcolor{black}{Marcel G. Clerc}}

\affiliation{\textcolor{black}{Departamento de F�sica, Facultad de Ciencias F�sicas
y Matem�ticas, Universidad de Chile, Casilla 487-3, Santiago, Chile.}}

\author{\textcolor{black}{Isabelle Robert-Philip}}

\affiliation{\textcolor{black}{Centre de Nanosciences et de Nanotechnologies,
CNRS, Univ. Paris-Sud, Universit� Paris-Saclay, C2N Marcoussis, 91460
Marcoussis, France}}

\author{\textcolor{black}{R�my Braive{*}}}

\affiliation{\textcolor{black}{Centre de Nanosciences et de Nanotechnologies,
CNRS, Univ. Paris-Sud, Universit� Paris-Saclay, C2N Marcoussis, 91460
Marcoussis, France}}

\affiliation{\textcolor{black}{Universit� Paris Diderot, Sorbonne Paris Cit�,
75207 Paris Cedex 13, France}}

\date{\textcolor{black}{\today}}
\begin{abstract}
\textcolor{black}{Stochastic resonance is a general phenomenon usually
observed in one-dimensional, amplitude modulated, bistable systems.We
show experimentally the emergence of phase stochastic resonance in
the bidimensional response of a forced nano-electromechanical membrane
by evidencing the enhancement of a weak phase modulated signal thanks
to the addition of phase noise. Based on a general forced Duffing
oscillator model, we demonstrate experimentally and theoretically
that phase noise acts multiplicatively inducing important physical
consequences. These results may open interesting prospects for phase
noise metrology or coherent signal transmission applications in nanomechanical
oscillators. Moreover, our approach, due to its general character,
may apply to various systems.}
\end{abstract}
\maketitle
\textcolor{black}{Stochastic resonance whereby a small signal gets
amplified resonantly by application of external noise has been introduced
originally in paleoclimatology \cite{Benzi,GammaitoniRMP98} to explain
the recurrence of ice ages and has then been observed in many other
areas including neurobiology \cite{Douglass,Bezrukov}, electronics
\cite{FauvePLA83}, mesoscopic physics \cite{Hibbs}, photonics \cite{McNamara,PhysRevE.61.157},
atomic physics \cite{PhysRevLett.85.1839} and more recently mechanics
\cite{BadzeyNat05,PhysRevA.79.031804,VenstraNatCom13,MonifiNP16}.
Implementation of stochastic resonances involves generally three ingredients
: (i) the existence of metastable states separated by an activation
energy, as in excitable or bistable nonlinear systems, (ii) a coherent
excitation, whose amplitude is however too weak to induce deterministic
hopping between the states, and (iii) stochastic processes inducing
random jumps over the potential barrier. In the classical picture
of a bistable system, this corresponds to the motion of a fictive
particle in a double-well potential periodically modulated in amplitude
by the signal and subjected to noise \cite{PhysRevLett.62.349}. When
an optimal level of noise is reached, the system's response power
spectrum displays a peak in the signal to noise ratio, unveiling the
stochastic resonance phenomenon. The resonance occurs as a 'bona-fide'
resonance in a frequency band around a signal frequency approximately
given by the time-matching condition \cite{GammaitoniPRL95,BarbayPRL00},
i.e. when the potential modulation period is twice the mean residence
time of the noise-driven particle. Experimental works on stochastic
resonance are almost exclusively using amplitude modulation going
along with additive amplitude noise or multiplicative amplitude noise
\cite{PhysRevE.49.4878,PhysRevE.61.940,WuJoMO07,WuPRA09,QiaoPRE16}.
In this case, it corresponds to a pure one dimensional effect. Few
studies take advantage of a bidimensional phase space by e.g. using
phase modulation and/or phase noise (i.e. phase random fluctuations
of input signal) \cite{doi:10.1021/nl9004546,Guerra}. Most of them
use amplitude noise to demonstrate amplitude stochastic resonance,
or introduce noise in the form of the response of a stochastic oscillator
\cite{Schimansky-Geier1990}. However in the latter scheme, neither
the noise nor the modulation are controlled, thus preventing to unveil
the specific roles of phase modulation and phase noise in stochastic
resonance.}

\textcolor{black}{In this Letter, stochastic resonance is implemented
in a nonlinear nanomechanical oscillator forced close to its resonant
frequency. It enables, in a bidimensionnal phase space, the implementation
of phase stochastic resonance observed simultaneously both on the
phase and amplitude response of the oscillator. It is here demonstrated
by achieving the stochastic enhancement of a phase modulated signal
by phase noise observed on the bidimensional response of the oscillator.
This opens new avenues for stochastic resonance in bidimensional systems
by allowing for instance stochastic amplification of mixed phase-amplitude
modulated signals by complex value noise. We highlight that the system's
response can be projected on any variable in phase space and that
the amplification depends on the chosen basis. Finally, we derive
a stochastic nonlinear amplitude equation for the forced stochastic
Duffing oscillator, which describes qualitatively well our system,
and show that phase noise acts multiplicatively inducing important
physical consequences. }

\begin{figure}[H]
\begin{centering}
\textcolor{black}{\includegraphics[width=1\columnwidth]{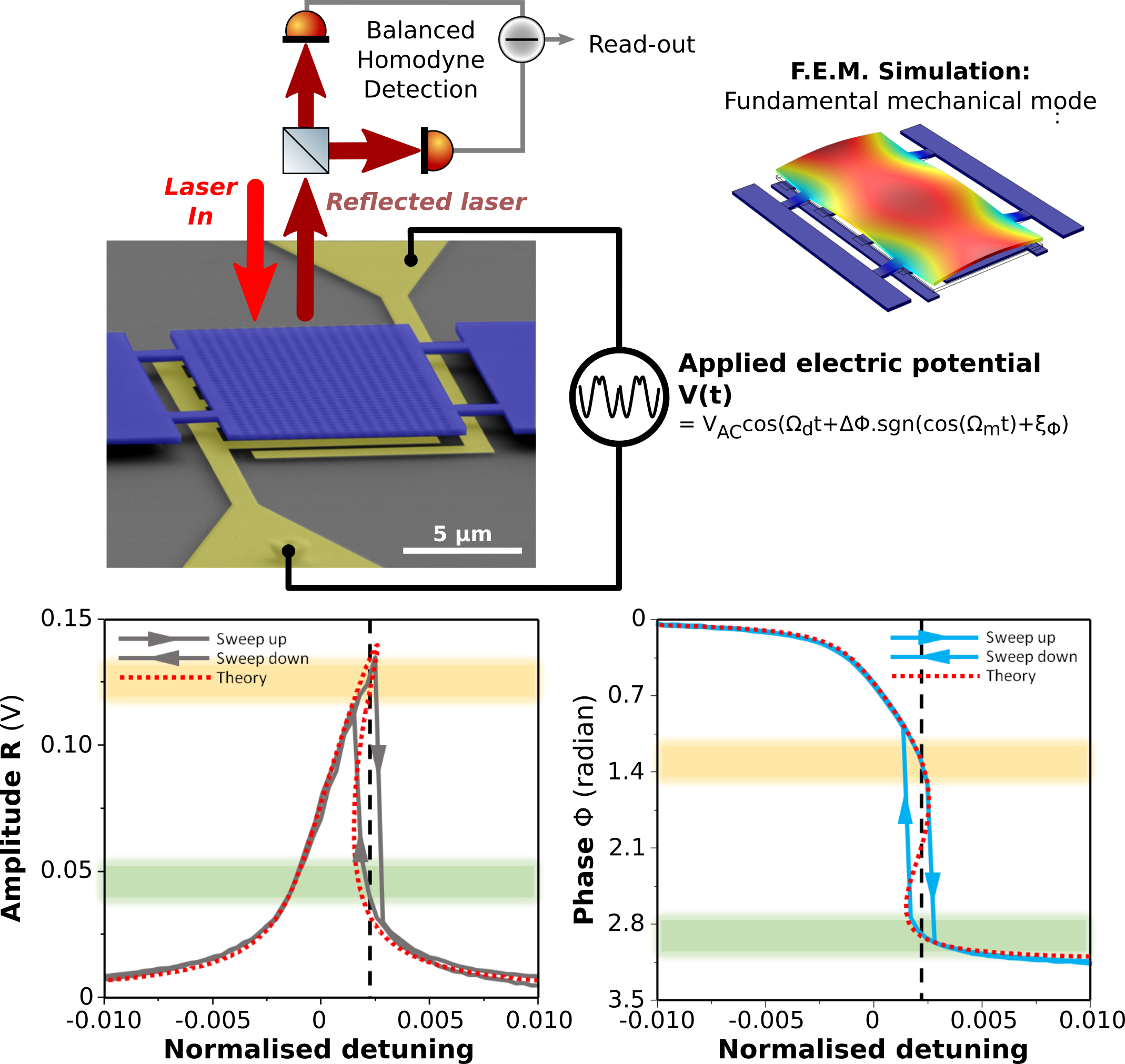} }
\par\end{centering}
\textcolor{black}{\caption{\textcolor{black}{(Top left) A scanning electron microscopic view
of the device shows the membrane (thickness of 260 nm and a 10$\times$20$\mu$m$^{2}$
surface) forming the mechanical oscillator (purple) and the interdigitated
electrodes (yellow) underneath, at a distance of about 400 nm. (Top
right) Finite Element Model (F.E.M.) simulation of the fundamental
mechanical mode under study with enhanced out-of-plane displacement
for clarity. (Bottom left) Amplitude $R$ and (bottom right) phase
$\Phi$ spectra of the driven oscillator response in a frequency sweep-up
and sweep-down experiment with $V_{AC}=9\,V$, $\Delta\phi=0$ and
$\xi_{RMS}=0$; the theoretical response is displayed in red dashed
lines in both spectra. The driving frequency $\nicefrac{\Omega_{d}}{2\pi}=2.824\,MHz$
(dashed lines) lies close to the linear fundamental mechanical resonance
at ${\color{red}{\color{black}\nicefrac{\Omega_{0}}{2\pi}=}}2.822\,MHz$
(i.e. zero normalised detuning). Normalized detuning is defined as
$\left(\Omega_{d}-\Omega_{0}\right)/\Omega_{0}$.}}
}

\textcolor{black}{\label{SetUp} }
\end{figure}

\textcolor{black}{The forced nanomechanical oscillator consists of
a suspended InP photonic crystal membrane which acts as a mirror in
one arm of an interferometer fed with an He-Ne laser. The membrane
is activated by underneath integrated interdigitated electrodes driven
by an AC-bias voltage $V\left(t\right)$ (see Fig. \ref{SetUp}).
This voltage induces an electrostatic force on the oscillator which
drives its out-of-plane motion as described in \cite{Chowdhury}.
The oscillator is placed in a vacuum chamber with a pressure of about
$10^{-4}$ mbar at room temperature. The phase $\Phi$ and the amplitude
modulus $R$ of the oscillator's motion are retrieved by use of a
balance homodyne detection. From the recorded time traces of $\Phi$
and $R$, we can reconstruct the polar plots with the two quadratures
$X=R\cos\left(\Phi\right)$ and $Y=R\sin\left(\Phi\right)$. }

\textcolor{black}{}

\textcolor{black}{The applied voltage, and therefore the applied electrostatic
force, is in the form of \cite{makles:tel-01290469}: }

\textcolor{black}{
\begin{equation}
V\left(t\right)=V_{AC}\cos\left[\Omega_{d}t+\Delta\phi\ \mathrm{sgn}\left(\cos\left(\Omega_{m}t\right)\right){\color{black}{\color{red}{\color{black}+\xi\left(t\right)}}}\right].
\end{equation}
}

\textcolor{black}{Here $V_{AC}$ is the amplitude of the applied voltage,
while $\Omega_{d}$ denotes the resonant driving frequency. A phase
modulation is added; it displays a square waveform described by the
sign function $sgn$, at frequency $\Omega_{m}$ and a phase deviation
of $\Delta\phi$. Gaussian phase noise $\xi(t)$ of zero-mean and
standard deviation $\xi_{RMS}$ (bandwidth $B_{\phi}=10\,kHz$ such
that $\Omega_{m}\ll B_{\phi}$) is also applied on the nonlinear dynamic
system. Under quasi-resonant forcing of the mechanical fundamental
mode, a hysteresis behavior becomes prominent for $V_{AC}>5\,V$ and
two stable fixed points co-exist in the bidimensional phase space
of the oscillator (Fig.~\ref{SetUp}-bottom left and right). In the
following, $V_{AC}$ is set to $9\,V$ in order to be deeply in the
bistable regime; the driving frequency is set inside the hysteresis
region at $\nicefrac{\Omega_{d}}{2\pi}=2.824\,MHz$ in order to get
equal probability of residence in each state (See Supplementary Information)
and the system is systematically initially prepared in its upper state.}

\begin{figure}
\begin{centering}
\textcolor{black}{\includegraphics[width=0.9\columnwidth]{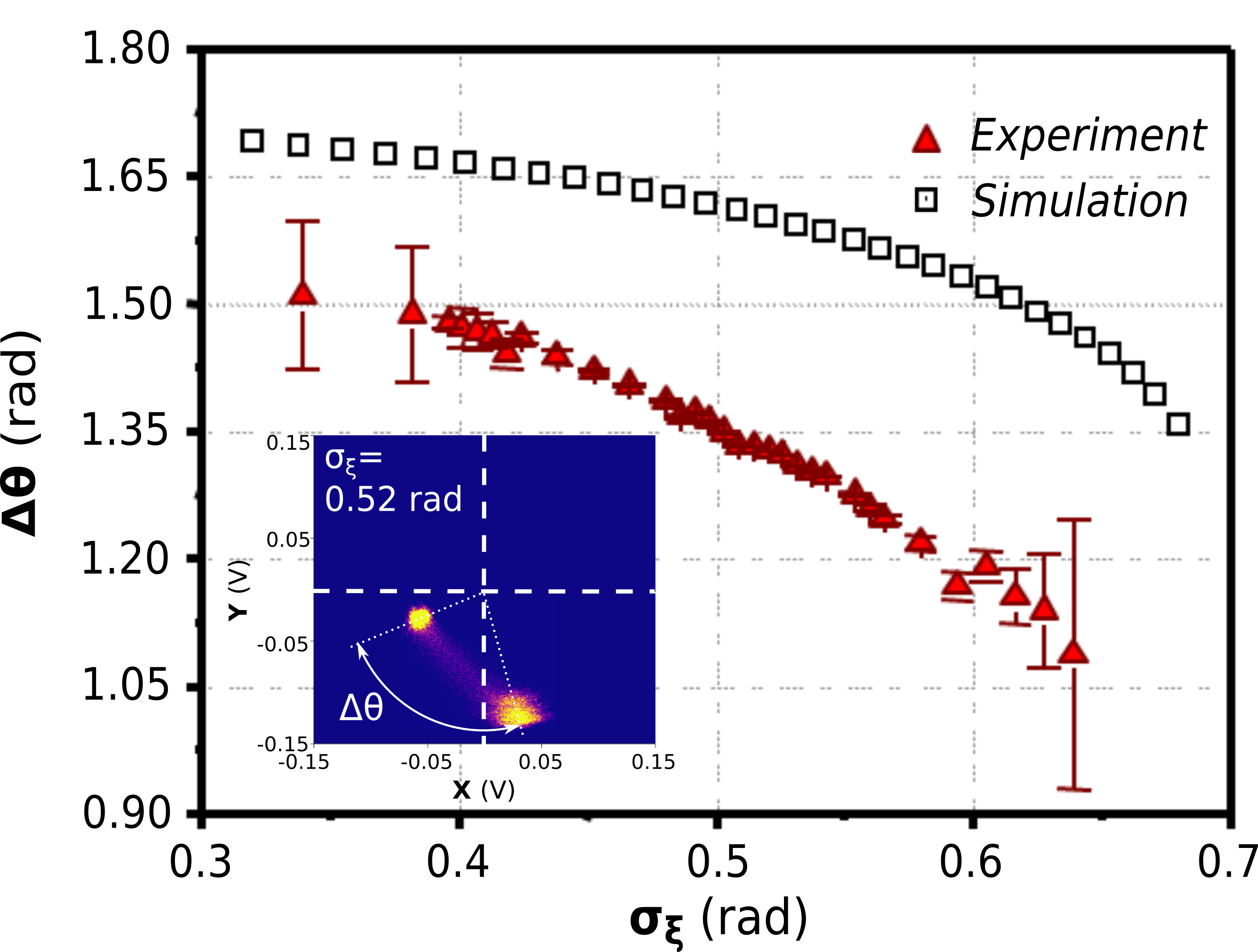} }
\par\end{centering}
\textcolor{black}{\caption{Simulated (open squares) and experimental (red triangles) polar angle
difference between the two stable states for increasing noise strengths.
Inset: Experimental polar plots with experimental values of $\xi_{RMS}=0.52\,rad$
and $\nicefrac{\Omega_{d}}{2\pi}=2.824\,MHz$.}
\label{Fig2} }
\end{figure}

\textcolor{black}{In the bistability regime, jumps between the two
stable states can be induced by applying a slow modulation ($\Omega_{m}\ll\Omega_{d}$)
with a sufficiently high phase deviation, phase noise strength or
both. These jumps are investigated by tracing the amplitude and phase
evolution of the fundamental mode with time and are also pictured
in the X-Y phase plane. In the case of pure phase modulation, the
system can transit or not from one state to the other depending on
the values of $\Omega_{m}$ and $\Delta\phi$. Beyond the cut-off
frequency $\nicefrac{\Omega_{m,c}}{2\pi}=1\,kHz$, which is directly
linked to the oscillator's line width of $0.9\,kHz$ \cite{Guerra},
the output signal is not synchronized with the input signal, in amplitude
or phase. For $\nicefrac{\Omega_{m}}{2\pi}=500\,Hz$, every jumps
in the input signal translate into a jump in the output signal for
$\Delta\phi>1.83\,rad$ (see Supplementary Information). Similarly,
in the case of pure noise-induced switching, the system starts to
transit between the two 2D states, in amplitude and phase as noise
strength increases. The occupancy between these two states becomes
equiprobable for values of $\xi_{RMS}$ close to $0.52\,rad$ in our
device. Such noise-induced transitions can also be quantified by the
Kramers rate $T_{K}=\nicefrac{1}{\tau_{K}}$ which is the inverse
of the average time required to cross over the barrier \cite{KRAMERS1940284}
and reaches a value close to $100\,Hz$ (see Supplementary Information).
Contrary to amplitude noise which amounts to additive noise, phase
noise acts here as a multiplicative noise. This feature is revealed
through the non-constant dependence of the phase difference $\Delta\theta$
between the two equilibria for increasing noise strengths (see Fig.~\ref{Fig2})
and is highlighted by the fourth term in the right hand side of Eq.~\ref{EqTheory2}.
At weak phase noise ($\xi_{RMS}<0.4\,rad$), uncertainties on the
phase difference are large because the probability of residence in
the lower state is weak ($<5\,\%$) and thus this state gets difficult
to observe. Conversely, at strong phase noise ($\xi_{RMS}>0.6\,rad$),
the probability of residence of the upper state reduces, and this
state is hardly observable.}

\textcolor{black}{}

\begin{figure}
\begin{centering}
\textcolor{black}{\includegraphics[width=1\columnwidth]{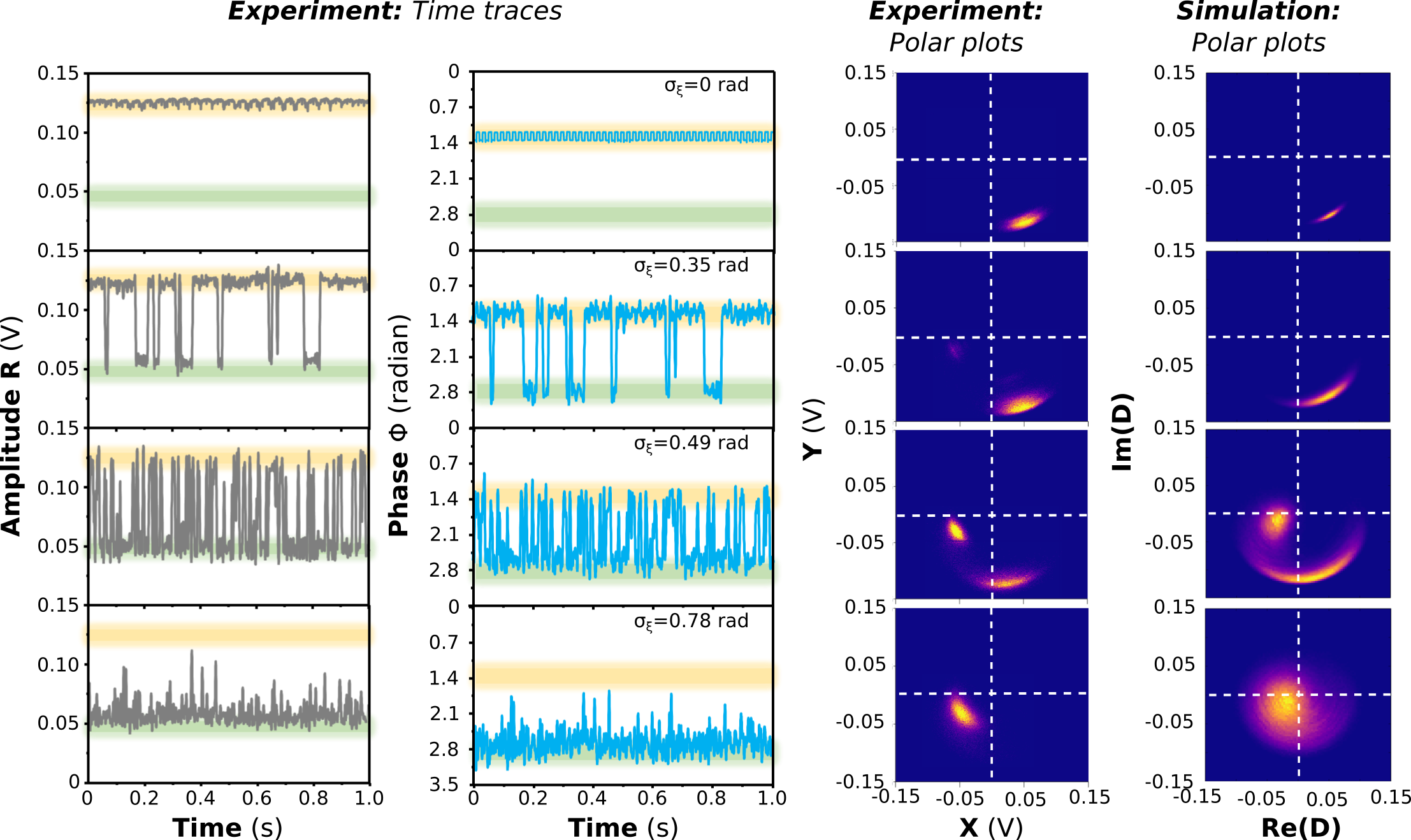} }
\par\end{centering}
\textcolor{black}{\caption{The response of the system, now driven by a force combining a weak
phase modulation and an increasing phase noise is shown. (From left
to right) Experimental time traces recorded on a timescale of $300\,s$
of the amplitude $R$ and phase $\Phi$ of the fundamental mode for
increasing noise strengths, with associated experimental and theoretical
polar plots. (From upper to lower lines) Evolution of these four panels
for increasing standard deviation $\xi_{RMS}$. \label{Fig3}}
}
\end{figure}

\textcolor{black}{The stochastic synchronization between the external
noise and the weak coherent signal that occurs in stochastic resonance,
takes place when the average waiting time between two noise-induced
interwell transitions ($T_{K}$) is comparable to half the period
of the periodic signal ($T_{\varOmega}=\nicefrac{2\pi}{\Omega_{m}}$).
In order to match this time-scale condition, modulation frequency
$\Omega_{m}$ in phase is set at $50\,Hz$. The deviation $\Delta\phi$
is also set to $0.09\,rad$ ($\ll1.56\,rad$, the hysteresis width
(Fig.~\ref{SetUp})), a far too weak value to let the system switch
periodically from one state to the other (see Fig.~\ref{Fig3} upper
line). When increasing the noise strength, occasional transitions
occur, weakly locked to the modulation signal. For $\xi_{RMS}=0.49\,rad$,
the transitions get stochastically synchronized with the modulation
(see Fig.~\ref{Fig3}). Further increasing the noise distorts the
hysteresis cycle and the system drops to its lower state.}

\textcolor{black}{}

\begin{figure*}
\begin{centering}
\textcolor{black}{\includegraphics[width=1\textwidth]{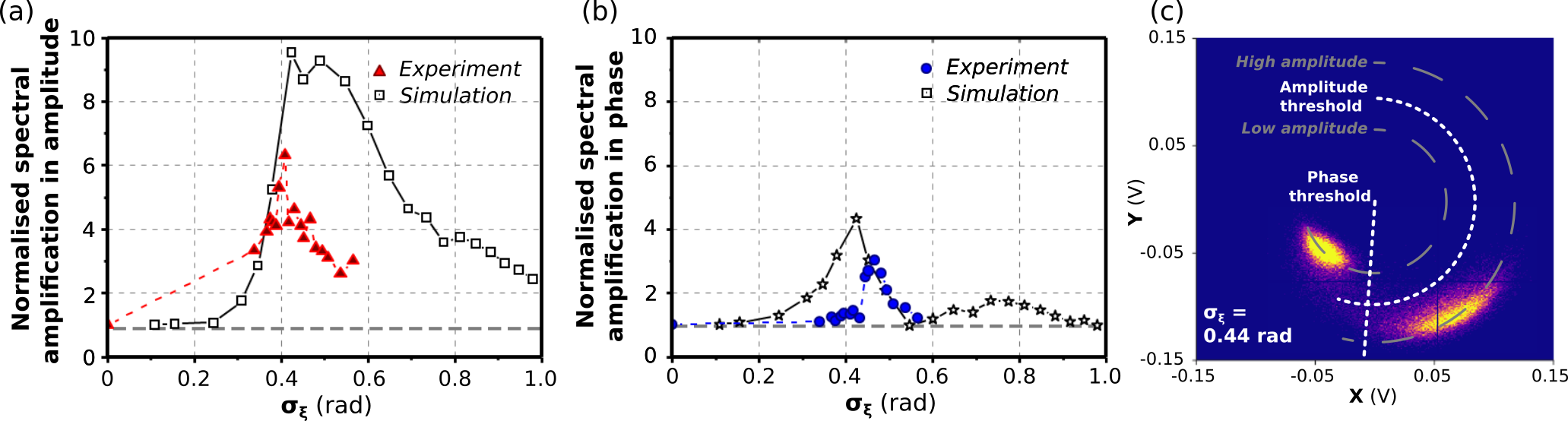} }
\par\end{centering}
\textcolor{black}{\caption{(a) Experimental (red triangles) and theoretical (open black squares)
spectral amplification in amplitude $R$ as a function of phase noise
strength. (b) Experimental (blue circles) and theoretical (open black
stars) spectral amplification in phase $\Phi$ as a function of phase
\textcolor{black}{noise} stren\textcolor{black}{gth. Theoretical curves
have been obtained by use of Eq.(\ref{EqTheory2}). (c) Experimental
polar plot for $\xi_{RMS}=0.44\,rad$ (maximum amplification), highlighting
the shape of the two stable points as well as the directions imprinted
by the modulation on the input phase or amplitude. The two dashed
grey lines are guide for the eyes indicating the two distinct amplitude
states and the dotted white line highlights the threshold. }}
\label{Fig4} }
\end{figure*}

\textcolor{black}{Quantification of achieved amplification relies
on a Discrete Fourier Transform (DFT) of the time traces. The spectral
power amplification is then given by the ratio between the strength
of the peak in the DFT at $\Omega_{m}$ for a given noise intensity
and its strength without added noise. For both variables, $R$ and
$\Phi$, evolution of the spectral amplification is observed as a
function of the phase noise strength and are plotted on Fig.~\ref{Fig4}-a
and b. It presents a bell-shaped maximum which reaches, for the amplitude
variable, a value up to ${\color{black}{\color{black}6.3}}$ and peaks
at $\xi_{RMS}=0.44\,rad$ (see Fig. \ref{Fig4}-a). This noise strength
is close to the one at which the system has a Kramer's rate of about
$100\,Hz$ with only noise applied. Under the same conditions, amplification
of the phase variable is also shown in Fig. \ref{Fig4}-b. It reaches
experimentally a value up to 3 for the same phase noise strength.
A double peak is clearly visible in the numerical spectral amplification
of the phase. The first peak is indeed attributed to the synchronized
hopping between the two metastable states, whereas the other peak
is due to an internal state resonance \cite{PhysRevE.62.299}. For
higher noise strength, the noise-induced effective detuning makes
a longer residence time in the lower state, and the Kramers rates
are not balanced anymore.}

\textcolor{black}{}

\textcolor{black}{To gain more insight into the observed dynamics,
we compare our results to theoretical and numerical predictions of
a stochastic amplitude equation. Fits of the experimental results
are obtained by modeling the nano-electromechanical oscillator by
a simple forced stochastic Duffing oscillator \cite{Duffing18} whose
dynamics can be described, in the limit of small injection and dissipation
of energy, by: 
\begin{multline}
\ddot{x}=-x-\epsilon\mu\dot{x}-\alpha x^{3}+\\
\epsilon^{3/2}F\cos((1+\epsilon\sigma)t+\epsilon\phi_{m}(t)+\epsilon\sqrt{\eta_{0}}\Delta W_{\phi})\label{eq:Duffing2}
\end{multline}
where $x(t)$ accounts for the displacement of the membrane and $\epsilon$
is a small control parameter ($\epsilon\ll1$). This parameter is
introduced to properly balance the scaling between the dissipation
and injection of energy in the system, and also control the frequency
detuning. The natural frequency has been rescaled to one ($\omega_{0}=1$),
$\mu\ll1$ is the damping coefficient that accounts for dissipation
of energy, $\alpha$ accounts for the nonlinear stiffness of the spring,
which is positive (negative) for soft (hard) spring \cite{Bogoliubov61}
and $F$ the strength of the driving. The near-resonant drive has
an angular frequency of $\omega_{d}=1+\sigma$, where $\sigma\ll1$
stands for the detuning between the drive and the natural resonant
frequency. The system is also subject to a slow phase modulation $\phi_{m}(t)$
($\dot{\phi}_{m}\ll\omega_{0}\phi_{m}$) and to a phase noise term
in the form of a Wiener process $\Delta W_{\phi}$ with Gaussian noise
strength $\eta_{0}$. In the conservative limit and for small displacements,
the system exhibits harmonic motion with a small arbitrary amplitude
$D$ such that $x(t)=Re[De^{it}]$. When considering the nonlinear
terms, dissipation and forcing, the displacement of the membrane response
can be approximated by \cite{Bogoliubov61,Kevorkian96}: 
\begin{eqnarray}
x(t) & = & \epsilon^{3/2}D(T=\epsilon t)e^{i[t+\epsilon(\sigma t+\phi_{m}(t)+\sqrt{\eta_{0}}\Delta W_{\phi})]}\nonumber \\
\  &  & +\frac{\alpha\epsilon^{9/2}}{8}D^{3}e^{i3[t+\epsilon(\sigma t+\phi_{m}(t)+\sqrt{\eta_{0}}\Delta W_{\phi})]}+c.c.+o(\epsilon^{5})\label{Ansatz}
\end{eqnarray}
where the envelope of the oscillations $D$ is promoted to a temporal
variable \cite{Bogoliubov61,Kevorkian96,NewellARoFM93}, $T$ accounts
for the slow temporal scale ($\dot{D}\simeq\epsilon D$ and $\ddot{D}\simeq\epsilon^{2}D$),
and the symbol $c.c.$ stands for complex conjugate. Introducing the
above ansatz in Eq. (\ref{eq:Duffing2}) to order $\epsilon^{3/2}$
and using the rules of calculus in stochastic normal form theory \cite{ClercPRE06}
one finds the stochastic nonlinear amplitude equation: 
\begin{equation}
\frac{dD}{dT}=-(\frac{\mu}{2}+i(\sigma+\frac{d\phi_{m}}{dT}+\sqrt{\eta_{0}}\xi))D+\frac{3i\alpha}{8}|D|^{2}D-\frac{iF}{2}\label{EqTheory2}
\end{equation}
where $\xi=d\Delta W_{\phi}/dT$ is a zero-mean and delta-correlated
white Gaussian noise term. Note that $\phi_{m}$ is a slow phase modulation,
that is, $d\phi_{m}/dt=\epsilon d\phi_{m}/dT$. To derive the above
model, we have considered ansatz (\ref{Ansatz}) as a change of variable.
Here, the Stratonovich prescription for noise has been adopted. Namely,
the stochastic term can induce a non-zero drift, $\left\langle \xi\left(T\right)D\left(T\right)\right\rangle \neq0$.
Note that even though Eq.(\ref{eq:Duffing2}) would give rise to additive
noise with time-dependent coefficients in a Fokker-Planck equation,
the reduced equation (Eq.(\ref{EqTheory2})) for the response amplitude
of the oscillations satisfies a stochastic differential equation with
multiplicative noise as a result of the stochastic normal-form derivation
\cite{ClercPRE06}.}

\textcolor{black}{Stochastic numerical simulations of Eq.(\ref{EqTheory2})
are performed with the help of the XMDS2 package \cite{DennisCPC13}.
We use the semi-implicit numerical scheme which converges to the Stratonovich
integral. The time step is kept fixed in the simulation and is chosen
to be $dT=0.1$. The slow phase modulation is sinusoidal with an amplitude
$\Delta\phi=5\times10^{-5}$ and an angular frequency $\Omega_{m}=2\pi/2\times10^{5}$.
The detuning is $\sigma=1.77\times10^{-3}$. The model reproduces
well the bistable response in amplitude and phase of our nano-electromechanical
oscillator (see Fig.~\ref{SetUp} bottom), as well as the temporal
evolution of the response in amplitude or phase, in the case of pure
phase modulation, pure phase noise (see Supplementary Information)
and stochastic resonance (see Fig.~\ref{Fig3} and \ref{Fig4}).
Moreover, multiplicative noise shall translate into a shift of the
operating point in the hysteresis and thus into an effective detuning
in Eq. (\ref{EqTheory2}) which reduces to $\sigma_{eff}=\sigma+\eta_{0}/2$.
Physically, this translates in a drift of the operating point for
increased noise strengths, a signature of the multiplicative nature
of the added noise, as observed in our experiment (see Fig.~\ref{Fig2}).
The measured $\Delta\theta$ is slightly smaller in the experiment
compared to theory presumably because of extra low-frequency noise
sources which are not taken into account in the model.}

\textcolor{black}{Stochastic resonance amplification of the modulated
signal is here limited by the relative orientation of the modulation
and of the minimal energy path between the two basins of attraction,
which is almost in a direct straight line (see Fig. \ref{Fig4}-c).
In the same frame, the added phase modulation shakes the upper state
preferentially in the azimuthal direction. These two orientations
being not parallel, higher amplification value can not be achieved
in this configuration. This reveals the importance of the modulation
format of the signal: optimal stochastic resonance would certainly
require a mixed amplitude-phase format to follow the minimal energy
path in the nanomechanical oscillator phase space. The distribution
of the two states in the phase plane gets also distorted: The system
switches between a symmetric branch (with a quasi-circular state in
the phase portrait) to an asymmetric branch (with an elongated state
in the polar plot). Such distortion is reminiscent to thermal noise
squeezing observed e.g. in parametrically-driven oscillators \cite{RugarPRL91,BriantEPJD03,SzorkovszkyPRL13,PontinPRL16}.}

\noindent \textcolor{black}{In conclusion, we have demonstrated phase
stochastic resonance with phase noise in a bidimensional nonlinear
oscillator consisting of a nano-electromechanical device. The applied
phase noise reveals to act as a multiplicative noise on the system
which introduces an effective detuning that plays a crucial role in
the residence probability asymmetry. The derived stochastic amplitude
equation (\ref{EqTheory2}) is a universal model that describes the
evolution of the envelope of the oscillations near a nonlinear resonance
and subjected simultaneously to phase noise and to a phase modulation.
That is, it applies to any nonlinear oscillator with such forcing
provided one makes use of a suitable nonlinear and periodic change
of variables in the initial equations that describe the system. Our
model applies to e.g. dispersive optical bistability that plays an
important role in nonlinear optical science \cite{TalknerPRA84} and
can thus shed a new light on coherent processes involving phase fluctuations
in these systems \cite{CasteelsPRA17}. Such stochastic resonance
obtained by the assistance of phase noise may also enable various
noise-aided applications, including signal transmission \cite{JungPRA91,PhysRevLett.85.3369}
in particular involving novel coherent schemes such as Phase Key Shifting
protocol, or metrology with improved detection in noise-floor limited
systems \cite{PhysRevE.61.940,PhysRevLett.102.080601,10.1371/journal.pone.0109534}. }

\paragraph*{\textcolor{black}{Acknowledgements}}

\textcolor{black}{This work is supported by the ``Agence Nationale
de la Recherche\textquotedblright{} programme MiNOToRe, the French
RENATECH network, the Marie Curie Innovative Training Networks (ITN)
cQOM and the European Union\textquoteright s Horizon 2020 research
and innovation program under grant agreement No 732894 (FET Proactive
HOT). .}

\textcolor{black}{\bibliographystyle{apsrev4-1}}

\bibliographystyle{apsrev4-1}

\end{document}